\documentclass[aip,amsmath,amssymb,reprint]{revtex4-1}
\usepackage{graphicx}
\usepackage{dcolumn}
\usepackage{bm}

\usepackage[utf8]{inputenc}
\usepackage[T1]{fontenc}
\usepackage{mathptmx}
\usepackage{etoolbox}

\usepackage{color}

\makeatletter
\def\@email#1#2{%
 \endgroup
 \patchcmd{\titleblock@produce}
  {\frontmatter@RRAPformat}
  {\frontmatter@RRAPformat{\produce@RRAP{*#1\href{mailto:#2}{#2}}}\frontmatter@RRAPformat}
  {}{}
}%

\newcommand*{\citen}[1]{%
	\begingroup
	\romannumeral-`\x 
	\setcitestyle{numbers}%
	\cite{#1}%
	\endgroup
}

\makeatother

\begin{document}
\title{Circular cumulant reductions for macroscopic dynamics of oscillator populations with non-Gaussian noise}

\author{Anastasiya V.\ Dolmatova}
\affiliation{Institute of Continuous Media Mechanics, UB RAS, Academician Korolev
Street 1, 614013 Perm, Russia}
\author{Irina V.\ Tyulkina}
\affiliation{Institute of Continuous Media Mechanics, UB RAS, Academician Korolev
Street 1, 614013 Perm, Russia}
\affiliation{Department of Control Theory, Nizhny Novgorod State University, Gagarin Avenue 23, 603022 Nizhny Novgorod, Russia}
\author{Denis S.\ Goldobin}
\affiliation{Institute of Continuous Media Mechanics, UB RAS, Academician Korolev
Street 1, 614013 Perm, Russia}
\affiliation{Department of Control Theory, Nizhny Novgorod State University, Gagarin Avenue 23, 603022 Nizhny Novgorod, Russia}
\affiliation{Department of Theoretical Physics, Perm State University, Bukirev
Street 15, 614990 Perm, Russia}
\email{denis.goldobin@gmail.com}
\date{\today}
\begin{abstract}
We employ the circular cumulant approach to construct a low dimensional description of the macroscopic dynamics of populations of phase oscillators (elements) subject to non-Gaussian white noise.
Two-cumulant reduction equations for $\alpha$-stable noises are derived.
The implementation of the approach is demonstrated for the case of the Kuramoto ensemble with non-Gaussian noise. The results of direct numerical simulation of the ensemble of $N=1500$ oscillators and the ``exact'' numerical solution for the fractional Fokker--Planck equation in the Fourier space are found to be in good agreement with the analytical solutions for two feasible circular cumulant model reductions.
We also illustrate that the two-cumulant model reduction is useful for studying the bifurcations of chimera states in hierarchical populations of coupled noisy phase oscillators.
\end{abstract}

\maketitle

\begin{quotation}
While the mathematical theory of non-Gaussian stable random variables was significantly advanced in 1920s, the practical interest to non-Gaussian statistics was hyped by the observation that failures of Gaussian (normal) statistics for financial data can be easily repaired by implementation of stable non-Gaussian statistics. Non-Gaussian fluctuations naturally and robustly emerge even in the most basic mathematical setups: ranging from viscous flows through the arrays of solid obstacles or porous media to the synchronization by common noise (which can also serve as a theoretical framework for generalized synchronization of chaotic systems). The puzzlement with surprisingly low dimensionally of the collective dynamics of many paradigmatic population models was elucidated by rigorous mathematical explanations: the Watanabe--Strogatz and Ott--Antonsen theories. While these theories do not allow for individual noises in elements, the dynamics of noisy populations is low dimensional even more frequently and robustly. The circular cumulant formalism allowed to generalize the Ott--Antonsen theory to nonideal setups and encompass the case of Gaussian noise. Employing this formalism to populations subject to non-Gaussian stable noises, one can construct a low dimensional model reduction of collective dynamics. We present derivation of this reduction and explain why it circumvents usual issues with truncations of the generalized Fokker--Planck equation for non-Gaussian noise.
\end{quotation}

\vspace{0.75cm}

\section{Introduction}
The research interest to non-Gaussian fluctuations is related to financial, biological, and physical systems~\cite{Mandelbrot-1963,Pietras-etal-2023,Pyragas2-2023,Chechkin-etal-2003,Toenjes-etal-2013,Toenjes-Pikovsky-2020,Kostin-etal-2023,Goldobin-Pikovsky-2005,Goldobin-Dolmatova-2019}. Except for the case of Cauchy (Lorentzian) fluctuations~\cite{Pietras-etal-2023,Toenjes-Pikovsky-2020,Pyragas2-2023,Kostin-etal-2023}, for which the Ott--Antonsen theory~\cite{Ott-Antonsen-2008,Ott-Antonsen-2009} is exact, a microscopic theoretical description of systems with such fluctuations is challenging. For populations of phase oscillators/elements, a new formalism of circular cumulants was recently introduced~\cite{Tyulkina-etal-2018,Goldobin-etal-2018,Goldobin-2019,Goldobin-Dolmatova-2019b} and allowed for progress beyond the limits of applicability of the Ott--Antonsen theory. This approach proved to be seminal for the case of Gaussian noise~\cite{Tyulkina-etal-2018,Goldobin-etal-2018,Goldobin-Dolmatova-2020}, in particular, for constructing the mathematical theory of the emergence of gamma oscillations in networks of quadratic integrate-and-fire neurons~\cite{Ratas-Pyragas-2019,Zheng-Kotani-Jimbo-2021,Goldobin-2021,diVolo-etal-2022}. In this paper we suggest the implementation of the circular cumulant formalism to the case of non-Gaussian fluctuations.

For $\delta$-correlated noises, by virtue of the generalized central limit theorem, any non-Gaussian noise will converge to an $\alpha$-stable one as we switch from a discrete time to the continuous time limit. In this paper, we consider the case of $\alpha$-stable sign-symmetric noises which are relevant for a broad class of physical and some financial applications.

The paper is organized as follows. In Sec.~\ref{ssec:CCR}, we derive circular cumulant equations for population of phase oscillators (elements) driven by independent $\delta$-correlated $\alpha$-stable noises. In Secs.~\ref{ssec:Kur} and \ref{ssec:wrap}, two options for the closures of the circular cumulant equations are considered for the Kuramoto ensemble with non-Gaussian noise and tested against the results of direct numerical simulation for a large but finite ensemble.
In Sec.~\ref{ssec:Abr}, the effect of $\alpha$-stable noise on bifurcations of chimera states in hierarchical populations of phase oscillators is studied.
In Sec.~\ref{sec:Meth} ``Methods,'' we provide a concise introduction into the statistics of $\delta$-correlated $\alpha$-stable noises and the derivation of the fractional Fokker--Planck equation for these noises. Conclusions are summarized in Sec.~\ref{sec:concl}.

\section{Results}\label{sec:res}
\subsection{Circular cumulant reduction for the population dynamics under non-Gaussian noise}\label{ssec:CCR}
We consider a population of sin-coupled phase elements (oscillators) subject to additive $\delta$-correlated noises $\sigma\xi_j(t)$
\begin{equation}
\dot\varphi_j=\omega(t)+\mathrm{Im}(h(t)e^{-i\varphi_j})+\sigma\xi_j(t)\,,
\label{eq110a}
\end{equation}
where $\xi_j(t)$ are independent non-Gaussian noises drawn from a so-called $\alpha$-stable distribution.
Parameter $0<\alpha\le2$: $\alpha=1$ for Cauchy noise, $\alpha=2$ for Gaussian noise; for $0<\alpha<2$, the fluctuation probability density possesses power-law tails $\propto1/|\xi_jdt|^{1+\alpha}$. In Sec.~\ref{sec:Meth} ``Methods'' below, we recall that a centered symmetrically-distributed $\alpha$-stable random variable is comprehensively determined by its characteristic function $F(v)=\exp(-|cv|^\alpha)$ ($c$ controls the width of the distribution).
One can derive a fractional Fokker--Planck equation governing the dynamics of the average probability density $w(\varphi,t)$ (e.g., see Refs.~\citen{Chechkin-etal-2003,Toenjes-etal-2013}). In the Fourier space, where
\[
w(\varphi,t)=\frac{1}{2\pi}\sum_{m=-\infty}^{+\infty}a_m(t)e^{-im\varphi},
\]
$a_0=1$ and $a_{-m}=a_m^\ast$, the Fokker--Planck equation reads
\begin{equation}
\dot{a}_m=im\omega a_m+mh\,a_{m-1}-mh^\ast a_{m+1} -(\sigma m)^\alpha a_m\,.
\label{eq113}
\end{equation}

In Sec.~\ref{sec:Meth} ``Methods,'' we provide a concise introduction to the $\alpha$-stable random variables and an explanation why, for $\delta$-correlated noises, the $\alpha$-stable noises are the only physically and mathematically meaningful non-Gaussian noises.

Noticeably, Eq.~(\ref{eq113}) highlights the issue with constructing an analog of the Fokker--Planck equation by means of cumulant representation of noise, such as Eq.~(\ref{eq108}) in the Methods section. For all physically relevant cases of a $\delta$-correlated noise, one must adopt $\alpha$-stable distributions with either $0<\alpha<2$ or $\alpha=2$ (the case of the standard Fokker--Planck equation). The representation of a singular function $(\sigma m)^\alpha$ for $0<\alpha<2$ by a series in $(\sigma m)$ never works; it fails for large $(\sigma m)$, which are associated with the formation of distribution discontinuities, $m\to\infty$, and for a weak noise, $(\sigma m)\to0$, as well.

The issue of formation of unphysical short-wave features in the probability distribution due to high-order spatial derivatives is circumvented in low-order circular cumulant reductions. Moreover, it, hopefully, should not emerge at all with Eq.~(\ref{eq113}). Unfortunately, for a non-integer $\alpha$, the derivation of general circular cumulant equations~\cite{Tyulkina-etal-2018} may be problematic and involves fractional derivatives. Let us deal with two-cumulant approximations~\cite{Goldobin-etal-2018,Goldobin-Dolmatova-2019} and Eq.~(\ref{eq113}). One can also handle here the case of a Cauchy distribution of natural frequencies
\[
G(\omega)=\frac{\gamma}{\pi\left(\gamma^2+(\omega-\omega_0)^2\right)},
\]
which yields equations for the Kuramoto--Daido order parameters from Eq.~(\ref{eq113}):
\begin{equation}
\dot{Z}_m=m\big((i\omega_0-\gamma)Z_m+hZ_{m-1}-h^\ast Z_{m+1}\big) +\dot\Phi_t^{(\xi)}(\sigma m)\,Z_m\,,
\label{eq114}
\end{equation}
where $Z_0=1$ and $\dot\Phi_t^{(\xi)}(\sigma m)=-(\sigma m)^\alpha$, which is {\it the time-derivative of the logarithm of the noise characteristic functional} introduced in Sec.~\ref{sec:Meth}.

Further, we introduce ``circular cumulants''~\cite{Tyulkina-etal-2018,Goldobin-etal-2018,Goldobin-Dolmatova-2019b} related to $Z_m$ via the generating function. The moment-generating function is
\begin{equation}
F(k,t)=\langle\exp(ke^{i\varphi})\rangle=\sum_{m=0}^{\infty}Z_m(t)\frac{k^m}{m!}\,,
\label{eqFkt}
\end{equation}
and its logarithm yields the circular cumulant-generating function
\begin{equation}
\Psi(k,t)=k\frac{\partial}{\partial k}\ln{F(k,t)}=\sum_{m=1}^{\infty}\kappa_mk^m.
\label{eqPSIkt}
\end{equation}
For the first three circular cumulants
\begin{align}
\kappa_1&=Z_1\,,
\nonumber\\
\kappa_2&=Z_2-Z_1^2,
\nonumber\\
\kappa_3&=\frac{Z_3-3Z_2Z_1+2Z_1^3}{2}\,,
\nonumber
\end{align}
one can write $\dot\kappa_1=\dot{Z}_1$, $\dot\kappa_2=\dot{Z}_2-2Z_1\dot{Z}_1$, etc.; therefore, the chain of the dynamics equations of circular cumulants reads
\begin{align}
\dot{Z}_1&=(i\omega_0-\gamma)Z_1 +h-h^\ast(Z_1^2+\kappa_2) +\Phi_{\sigma}Z_1\,,
\label{eq115}
\\[5pt]
\dot{\kappa}_2&=2(i\omega_0-\gamma)\kappa_2 -4h^\ast(\kappa_3+Z_1\kappa_2)
\nonumber\\
&\qquad\qquad\qquad\qquad
+\Phi_{2\sigma}\kappa_2+\left(\Phi_{2\sigma}-2\Phi_{\sigma}\right)Z_1^2\,,
\label{eq116}
\\[5pt]
\dot{\kappa}_3&=3(i\omega_0-\gamma)\kappa_3 -h^\ast(9\kappa_4+6Z_1\kappa_3+3\kappa_2^2) +\Phi_{3\sigma}\kappa_3
\nonumber\\
&
+\frac32\left(\Phi_{3\sigma}-\Phi_{2\sigma}-\Phi_{\sigma}\right)\kappa_2Z_1 +\frac{\Phi_{3\sigma}-3\Phi_{2\sigma}+3\Phi_{\sigma}}{2}Z_1^3\,,
\label{eq117}
\end{align}
where $\Phi_{m\sigma}:=\dot\Phi_t^{(\xi)}(m\sigma)$.
In Sec.~\ref{ssec:CCChain}, the infinite chain of the circular cumulant equations is derived; for a noninteger $\alpha$, the higher order equations cannot be written in an explicit regular form.

With $\dot\Phi_t^{(\xi)}(\sigma m)=-(\sigma m)^\alpha$, Eqs.~(\ref{eq115})--(\ref{eq116}) read
\begin{align}
&
\dot{Z}_1=(i\omega_0-\gamma)Z_1 +h-h^\ast(Z_1^2+\kappa_2) -\sigma^\alpha Z_1\,,
\label{eq118}
\\[5pt]
&
\dot{\kappa}_2=2(i\omega_0-\gamma)\kappa_2 -4h^\ast(\kappa_3+Z_1\kappa_2)
\nonumber\\
&\qquad\qquad\qquad\qquad
-\sigma^\alpha\left(2^\alpha\kappa_2 +(2^\alpha-2)Z_1^2\right)\,.
\label{eq119}
\end{align}

For $\alpha=1$ (Cauchy noise) or $\sigma=0$ (noise-free), the equation chain~(\ref{eq115})--(\ref{eq117}) admits solution $Z_1\ne0$, $\kappa_{m\ge2}=0$, where the dynamics of $Z_1$ is governed solely by Eq.~(\ref{eq115}); moreover, this solution is attracting~\cite{Ott-Antonsen-2009} for $\gamma>0$. Hence, the circular cumulant representation has potential to be a fruitful framework for studying nonideal situations, where higher $\kappa_m$ are nonzero but small.

For $\sigma>0$ and $\alpha\ne1$, Eqs.~(\ref{eq116}) and (\ref{eq117}) yield
\begin{align}
\kappa_2&\propto\sigma^\alpha(2^{\alpha-1}-1)Z_1^2\,,
\nonumber\\
\kappa_3&\propto\sigma^\alpha\frac{3^{\alpha-1}-2^\alpha+1}{2}Z_1^3
\nonumber\\
&\quad+\sigma^\alpha\frac{3^\alpha-2^\alpha-1}{2}\kappa_2Z_1 +\mathcal{O}\big(\kappa_2^2\big)\,.
\nonumber
\end{align}
Here several different smallness hierarchies can emerge:
\\
(i)~for a weak collective mode $|Z_1|\ll1$, circular cumulants $\kappa_2\propto Z_1^2$, $\kappa_3\propto Z_1^3$, \dots $\kappa_m\propto Z_1^m$;
\\
(ii)~for weak noise $\sigma\ll1$, circular cumulants $\kappa_2\propto\sigma^\alpha(2^{\alpha-1}-1)Z_1^2$ and $\kappa_3\propto0.5\sigma^\alpha(3^{\alpha-1}-2^\alpha+1)Z_1^3$, whence we can estimate the order of magnitude of the ratio
$\kappa_3/\kappa_2\sim C_{\sigma\to0}(\alpha)Z_1$ with coefficient $C_{\sigma\to0}(\alpha)=(3^{\alpha-1}-2^\alpha+1)/(2^{\alpha}-2)$;
\\
(iii)~for nearly-Cauchy noise with $\alpha=1+\alpha_1$, $|\alpha_1|\ll1$, one can calculate $2^{\alpha_1}-1=e^{\alpha_1\ln{2}}-1=\alpha_1\ln{2}+0.5(\alpha_1\ln{2})^2+\dots$ and
$\kappa_2\propto\alpha_1\sigma^{1+\alpha_1}Z_1^2\ln{2}$, $\kappa_3\propto\alpha_1\sigma^{1+\alpha_1}Z_1^3\ln(\sqrt{3}/2)$, whence $\kappa_3/\kappa_2\sim-0.2Z_1$.
\\
In case~(i), we have an obvious decaying geometric progression of $\kappa_m$. In case~(ii), the second circular cumulant makes a correction to the exact no-noise solution, but $\kappa_3$ is technically of the same order of smallness. However, firstly, $\kappa_3$ makes an indirect impact on the dynamics of $Z_1$, only via the dynamics of $\kappa_2$, which is immediately present in Eq.~(\ref{eq115}) for $Z_1$; secondly, $|C_{\sigma\to0}(\alpha)|$ is smaller than $1/3$ for all $\alpha$ and turns $0$ specifically for $\alpha=2$ (Gaussian noise), while $|Z_1|\le1$. Hence, $\kappa_3$ can be expected to be often of minor significance against the background of the $\kappa_2$-correction. In case~(iii), similarly to case~(ii), the $\kappa_3$-correction has the same order of smallness as the $\kappa_2$-correction, but can be expected to be much weaker. In summary, one can adopt an approximate closure $\kappa_3=0$ as a result of rigorous asymptotic expansion in some cases or as a rough approximation in other circumstances.

Practically, we observe self-organized hierarchies of smallness of $\kappa_m$ even without obvious small parameters in the system~\cite{Goldobin-2019,Goldobin-Dolmatova-2019b}, which makes the approximation $\kappa_3=0$ even more viable than one could expect from a formal ``skeptical'' analysis of the chain of equations for the hierarchy of smallness. The accuracy of this approximation was assessed and validated for all numerical results we present.

It is instructive to elucidate the geometric interpretation of the two first circular cumulants for a variable on the circumference via analogy with the variable on the infinite line~\cite{Goldobin-Dolmatova-2019b}.
For a variable on the line, two first conventional cumulants are real-valued and give the location of the center of the distribution and its width; the third and forth conventional cumulants quantify skewness and kurtosis, respectively.
For a variable on the circumference, the first circular cumulant (the Kuramoto order parameter) is complex-valued and quantifies the location of the distribution center and its width with $\arg{Z_1}$ and $\ln(1/|Z_1|)$, respectively. The second circular cumulant quantifies the distribution asymmetry with $(\arg\kappa_2-2\arg{Z_1})$ and the deformation of tails with $|\kappa_2|$, which are analogs of skewness and kurtosis, respectively.
Thus, for a variable on the line, the first four conventional cumulants are given by four real-valued numbers; for a variable of the circumference, the same amount of information comes with the absolute values and the arguments of the first two circular cumulants. The third circular cumulant provides characterization beyond the analogs of skewness and kurtosis.

\subsection{Numerical simulation of the Kuramoto model}\label{ssec:Kur}
We can test Eqs.~(\ref{eq113}) and (\ref{eq118})--(\ref{eq119}) with the Kuramoto model~\cite{Kuramoto-1975,Kuramoto-1984} endowed with intrinsic non-Gaussian noises. We employ the following protocol: (i)~to simulate a finite ensemble, (ii)~to solve Eqs.~(\ref{eq114}), (iii)~to solve Eqs.~(\ref{eq118})--(\ref{eq119}).

For the Kuramoto ensemble with noise
\begin{equation}
\dot\varphi_j=\omega_j+\varepsilon\sum_{m=1}^{N}\sin(\varphi_m-\varphi_j)+\sigma\xi_j(t)\,,
\quad j=1,...,N,
\label{eq201}
\end{equation}
one finds the shape~(\ref{eq110a}) with
\[
h(t)=\frac{\varepsilon Z_1}{2}\,.
\]
In the thermodynamic limit $N\to\infty$, Eqs.~(\ref{eq118})--(\ref{eq119}) with $\kappa_3=0$ yield a uniformly rotating solution ($\dot{Z}_m=im\omega_0Z_m$ or, equivalently, $\dot{\kappa}_m=im\omega_0\kappa_m$) with
\[
\kappa_2=\frac{(2-2^\alpha)\sigma^\alpha Z_1^2}{2\gamma+(2\sigma)^\alpha+2\varepsilon|Z_1|^2}
\]
and
\begin{align}
|Z_1|^2=\frac12-\frac{3(\gamma+\sigma^\alpha)}{2\varepsilon}
+\left(\frac14-\frac{\gamma+(3-2^{\alpha})\sigma^\alpha}{2\varepsilon}\right.
\nonumber\\
\left.{}+\frac{(\gamma+\sigma^\alpha)\left(\gamma+(9-2^{\alpha+2})\sigma^\alpha\right)}{4\varepsilon^2}\right)^\frac12.
\label{eq202}
\end{align}
Alternatively, discarding the corrections to the Ott--Antonsen theory~\cite{Ott-Antonsen-2008,Ott-Antonsen-2009,Goldobin-Dolmatova-2019} by setting $\kappa_2=\kappa_3=0$, one finds
\begin{equation}
|Z_1|^2=1-\frac{2(\gamma+\sigma^\alpha)}{\varepsilon}\,.
\label{eq203}
\end{equation}

From Eqs.~(\ref{eq115})--(\ref{eq117}), one can see that the linear stability threshold of the splash state of system~(\ref{eq114}), i.e.\ the Kuramoto transition point, is independent of higher circular cumulants $\kappa_2$, $\kappa_3$, etc. It can be calculated from (\ref{eq118}) with $\kappa_2=0$:
\[
\varepsilon_\mathrm{cr}=2(\gamma+\sigma^\alpha)\,.
\]
Solution~(\ref{eq202}) indicates that the bifurcation is always supercritical: there is a unique solution for $\varepsilon>\varepsilon_\mathrm{cr}$ and no solutions below the threshold.

For generation of $\alpha$-stable random variables, we employ the solution proposed in Ref.~\citen{Chambers-etal-1976} based on an integral formula from monograph~\citen{Zolotarev-1986}, as it was implemented in Ref.~\citen{Misiorek-Weron-2012}.

\begin{figure}[!t]
\center{
\includegraphics[width=0.47\textwidth]%
 {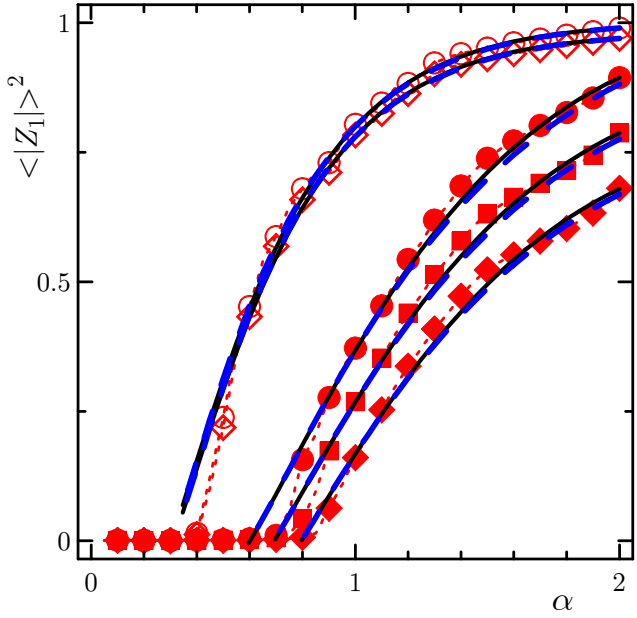}
}
\caption{The results of direct numerical simulation of $N=1500$ Kuramoto oscillators~(\ref{eq201}) are plotted with red symbols for $\varepsilon=1$: $(\sigma^2,\gamma)=(0.01,0)$ (open circles), $(0.01,0.01)$ (open diamonds),  $(0.1,0)$ (solid circles), $(0.1,0.05)$ (solid squares), $(0.1,0.1)$ (solid diamonds). Blue dashed lines: two-cumulant approximation~(\ref{eq202}), black solid lines: the rotating numerical solution ($\dot{Z}_m=im\omega_0Z_m$) of equation system~(\ref{eq114}) with 200 modes $Z_m$.
}
  \label{fig1}
\end{figure}

In Figs.~\ref{fig1} and \ref{fig2} one can see that the two-cumulant reduction provides a reasonably good approximation for the fractional Fokker--Planck equation~(\ref{eq114}) written in the Fourier space. The solutions heavily deviate from the Ott--Antonsen reduction~(\ref{eq203}) (not shown) except, obviously, for the case of the Cauchy noise, $\alpha=1$, where it is exact~\cite{Toenjes-Pikovsky-2020,Pietras-etal-2023,Pyragas2-2023}. The thermodynamic limit $N\to\infty$ solutions are also compared to the results of numerical simulation for a finite population (Fig.~\ref{fig1}). One can notice a stronger deviation of the finite population results for noninteger $\alpha$; this is presumably due to a poorer statistical convergence of the L\'evy flights.

\subsection{Wrapped distribution closures of the circular cumulant equation chain}\label{ssec:wrap}
Noise $\xi_j(t)$ creates fluctuations of $\varphi_j$ in an essentially nonlinear way and, generally, the distribution of $\varphi_j$ will deviate from an $\alpha$-stable one even for high degree of synchrony of the population. However, we use two independent order parameters $Z_1$ and $\kappa_2$ for the characterization of the population distribution and adopt an approximate closure for $\kappa_3$, which makes a nearly negligible impact on the dynamics of the former two. This suggests that one can also consider an option of a non-rigorous closure for $\kappa_3$ inspired by the properties of $\alpha$-stable distributions.

For a wrapped $\alpha$-stable distribution, the order parameters $\langle{e^{im\varphi}}\rangle$ equal to the values of the characteristic function $F(m;\alpha,\beta,\sigma,\mu)$ (\ref{eq105}) at integer $m$. For $\beta=0$,
\[
Z_m=e^{im\mu-\sigma^\alpha m^\alpha}.
\]
Circular cumulants $\kappa_2\propto\sigma^\alpha$ and $\kappa_3\propto\sigma^\alpha$, but not $\kappa_3\propto(\sigma^\alpha)^2$. Hence, a closure with the hierarchy $\kappa_{m+1}\kappa_{m-1}\sim\kappa_m^2$ does not work. Instead,
\begin{align}
\kappa_2&=-(2^\alpha-2)e^{i2\mu}\sigma^\alpha+\mathcal{O}(\sigma^{2\alpha})\,,
\nonumber\\
\kappa_3&=-\frac{3^\alpha-3\times2^\alpha+3}{2}e^{i3\mu}\sigma^\alpha+\mathcal{O}(\sigma^{2\alpha})\,;
\nonumber
\end{align}
therefore,
\begin{equation}
\kappa_3=\frac{3^\alpha-3\times2^\alpha+3}{2^{\alpha+1}-2^2}Z_1\kappa_2+\mathcal{O}(\sigma^{2\alpha})\,.
\label{eq204}
\end{equation}

Notice, this is in accordance with Eq.~(\ref{eq117}), where the coefficient of the last term vanishes only for $\alpha=1$, $2$; this term brakes the hierarchy $\kappa_m\sim(\sigma^\alpha)^{m-1}$. With closure~(\ref{eq204}), one can consider an alternative version of Eq.~(\ref{eq119}):
\begin{align}
&
\dot{\kappa}_2=2(i\omega_0-\gamma)\kappa_2 -2h^\ast\frac{3^\alpha-2^\alpha-1}{2^{\alpha}-2}Z_1\kappa_2
\nonumber\\
&\qquad\qquad\qquad\qquad
-\sigma^\alpha\left(2^\alpha\kappa_2 +(2^\alpha-2)Z_1^2\right)\,.
\label{eq205}
\end{align}

The rotating solution of system~(\ref{eq118}),\,(\ref{eq205}):
\begin{align}
  \kappa_2 & =-\frac{(2^\alpha-2)\sigma^\alpha Z_1^2}{2\gamma+(2\sigma)^\alpha +\varepsilon\frac{3^\alpha-2^\alpha-1}{2^\alpha-2}|Z_1|^2}\,,
\label{eq206}\\
|Z_1|^2&=\frac{1}{2}-\frac{(3^\alpha-3)(\gamma+\sigma^\alpha)}{(3^\alpha-2^\alpha-1)\varepsilon}
\nonumber\\
&\quad
+\Bigg(\left(\frac{1}{2}-\frac{(3^\alpha-3)(\gamma+\sigma^\alpha)} {(3^\alpha-2^\alpha-1)\varepsilon}\right)^2
\nonumber\\
&\quad
+\frac{(2^\alpha-2)(\varepsilon-2\gamma-2\sigma^\alpha)\left(2\gamma+(2\sigma)^\alpha\right)} {(3^\alpha-2^\alpha-1)\varepsilon^2}\Bigg)^\frac{1}{2}.
\label{eq207}
\end{align}
Solution~(\ref{eq207}) also corresponds to a supercritical bifurcation.

\begin{figure}[!t]
\center{
\includegraphics[width=0.47\textwidth]%
 {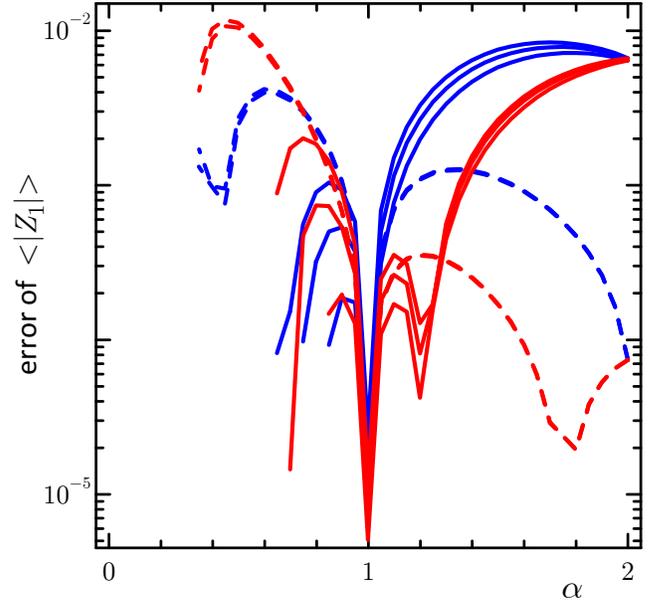}
}
\caption{The accuracy of two-cumulant approximations is calculated as the difference between the  approximation and the rotating numerical solution of equation system~(\ref{eq114}) with 200 modes $Z_m$. Blue lines: approximation~(\ref{eq202}) with $\kappa_3$ neglected, red lines: approximation~(\ref{eq207}) with closure~(\ref{eq204}) for $\kappa_3$ based on an $\alpha$-stable distribution approximation for the oscillator state fluctuations; solid: $\sigma^2=0.1$, $\gamma=0$, $0.05$, and $0.1$; dashed: $\sigma^2=0.01$, $\gamma=0$ and $0.01$.
}
  \label{fig2}
\end{figure}

In Fig.~\ref{fig2}, one can see that this closure provides a noticeable accuracy gain for $1<\alpha<2$, while for $\alpha<1$ the accuracy of two reductions is practically similar. The observed accuracy gain cannot be qualified as a uniform increase of the accuracy order with respect to small parameters of the system.

\subsection{Bifurcations of chimera states in hierarchical populations of coupled oscillators}\label{ssec:Abr}
Next to the threshold of the Kuramoto transition we studied in Secs.~\ref{ssec:Kur}, \ref{ssec:wrap} the smallness of $|Z_1|$ generates a hierarchy of smallness of $\kappa_m$, which creates conditions for a rigorous circular cumulant expansion. However, generally one does not have such obvious smallness hierarchy of $\kappa_m$ at a bifurcation point; moreover, marginal stability modes associated with the bifurcation favor the accumulation of the approximation error and create a challenge for approximate model reductions. In this section we illustrate a successful employment of the two-cumulant model reduction to the analysis of the effect of $\alpha$-stable noise on the bifurcations of the chimera states in hierarchical populations of coupled oscillators~\cite{Abrams-etal-2008,Pikovsky-Rosenblum-2008}.

The Ott--Antonsen (OA) theory~\cite{Ott-Antonsen-2008}, which practically amounts to a one-circular-cumulant reduction~\cite{Tyulkina-etal-2018,Goldobin-Dolmatova-2019b}, allowed Abrams {\it et al.}~\cite{Abrams-etal-2008} to conduct a detailed analytical study of chimera states in a hierarchical population of {\it noise-free} coupled identical Kuramoto oscillators. We consider the effect of additive $\alpha$-stable noise on this population:
\begin{align}
\dot\varphi_j&=\omega+\frac{1+A}{2N}\sum_{l=1}^N\sin(\varphi_l-\varphi_j-\vartheta)
\nonumber\\
&\qquad
+\frac{1-A}{2N}\sum_{l=1}^N\sin(\psi_l-\varphi_j-\vartheta)+\sigma\xi_{1j}(t)\,,
\nonumber\\
\dot\psi_j&=\omega+\frac{1+A}{2N}\sum_{l=1}^N\sin(\psi_l-\psi_j-\vartheta)
\nonumber\\
&\qquad
+\frac{1-A}{2N}\sum_{l=1}^N\sin(\varphi_l-\psi_j-\vartheta)+\sigma\xi_{2j}(t)\,.
\nonumber
\end{align}
Here two symmetric subpopulations of the same size $N$ are coupled with the coupling phase shift $\vartheta$ ($\vartheta=0$ for a diffusive coupling and $\pi/2$ for a reactive coupling), $\omega$ is the natural frequency of oscillators, $A$ determines the difference between intra- and interpopulation interactions, $\delta$-correlated noises $\xi_{sj}$ are independent.
In the thermodynamic limit $N\to\infty$, the macroscopic population dynamics is governed by equation chain~(\ref{eq113}) for the order parameters $Z_m$ and $Y_m$ of subpopulations $\{\varphi_j\}$ and $\{\psi_j\}$, respectively:
\begin{subequations}
\label{eqD01}
\begin{align}
\dot{Z}_m&=m\left(i\omega Z_m+H_1Z_{m-1}-H_1^\ast Z_{m+1}\right) -(\sigma m)^\alpha Z_m\,,
\label{eqD01Z}\\
\dot{Y}_m&=m\left(i\omega Y_m+H_2Y_{m-1}-H_2^\ast Y_{m+1}\right) -(\sigma m)^\alpha Y_m\,,
\label{eqD01Y}
\end{align}
\end{subequations}
where $H_{1,2}=0.25\left[(1\pm A)Z_1+(1\mp A)Y_1\right]e^{-i\vartheta}$ and $Z_0=Y_0=1$.
The two-cumulant reduction (\ref{eq118})--(\ref{eq119}) with neglected third circular cumulants for coupled equation chains~(\ref{eqD01}) reads
\begin{subequations}
\label{eqD02}
\begin{align}
\dot{Z}_1&=i\omega Z_1 +H_1-H_1^\ast\left(Z_1^2+\kappa_2\right) -\sigma^\alpha Z_1\,,
\label{eqD02a}
\\
\dot{\kappa}_2&=\left(2i\omega-4H_1^\ast Z_1 -(2\sigma)^\alpha\right)\kappa_2
-\left(2^\alpha-2\right)\sigma^\alpha Z_1^2\,,
\label{eqD02b}
\\
\dot{Y}_1&=i\omega Y_1 +H_2-H_2^\ast\left(Y_1^2+\chi_2\right) -\sigma^\alpha Y_1\,,
\label{eqD02c}
\\
\dot{\chi}_2&=\left(2i\omega-4H_2^\ast Y_1 -(2\sigma)^\alpha\right)\chi_2
-\left(2^\alpha-2\right)\sigma^\alpha Y_1^2\,,
\label{eqD02d}
\end{align}
\end{subequations}
where $\kappa_2$ and $\chi_2$ are the second circular cumulants for the first and second subpopulations, respectively.

\begin{figure}[!t]
\center{
\includegraphics[width=0.47\textwidth]%
 {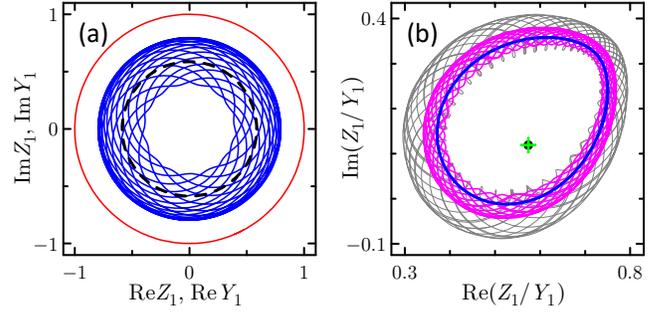}
}
\caption{Chimera state trajectories of hierarchical population~(\ref{eqD01}) are plotted for $\vartheta=\pi/2-0.15$ and no noise, $\sigma=0$. The red line forming the unit-radius circle in panel~(a) is the trajectory $Y_1$ of the synchronized subpopulation, $|Y_1|=1$. In panel~(b) the trajectories of the order parameter $Z_1$ of the partially synchronized subpopulation are plotted in the coordinate frame corotating with the synchronous subpopulation. The ``stable'' chimera existing at $A=0.28$ is seen a fixed point marked by the plus sign in the corotating frame, panel (b), and a rotating solution $Z_1(t)$ plotted with the black dashed line in panel~(a). For $A=0.31$, the ``breathing'' chimera (blue line) is a limit cycle in the rotating frame of panel~(b) and a torus in the original variables of panel~(a).
The grey and magenta lines in panel~(b) are the trajectories which start with different initial conditions away from the Ott--Antonsen manifold and oscillate around the breathing chimera cycle at $A=0.31$.
}
  \label{fig3}
\end{figure}
\begin{figure}[!t]
\center{
\includegraphics[width=0.47\textwidth]%
 {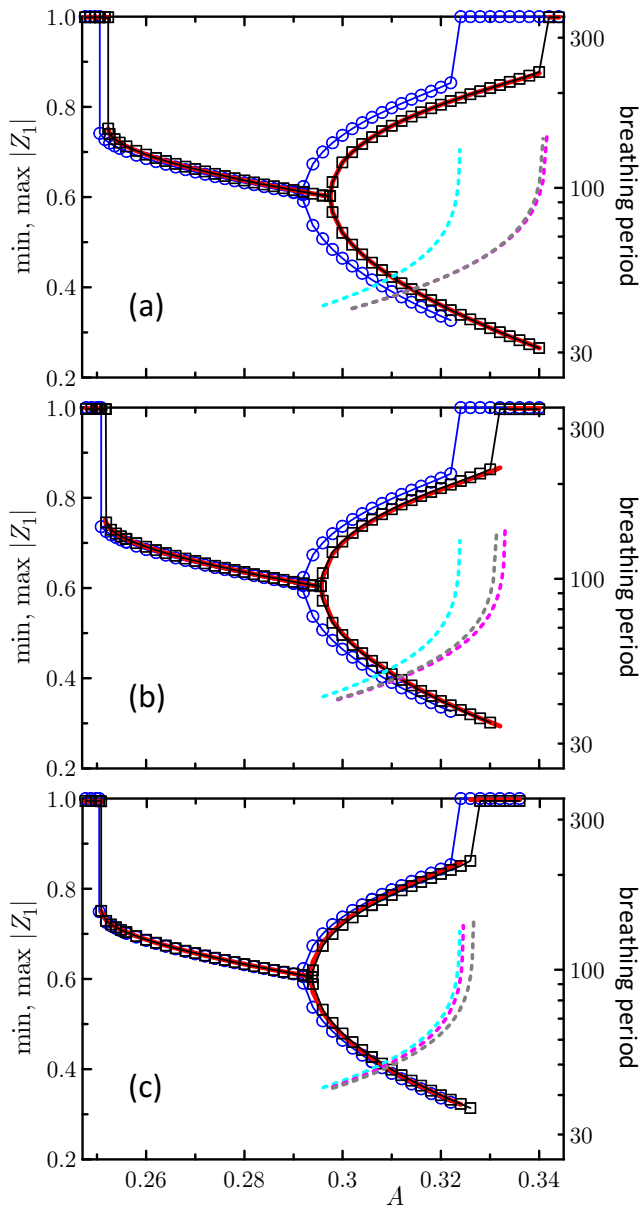}
}
\caption{Bifurcation diagrams of chimera states are plotted for coupling phase shift $\vartheta=\pi/2-0.15$, noise intensity $\sigma^\alpha=4\times10^{-4}$ and $\alpha=2$ (a), $1.5$ (b), and $0.5$ (c). The order parameter $|Z_1|$ is plotted for the noise-free system (blue circles), for equation chain~(\ref{eqD01}) with $200$ modes (black squares), for two-cumulant model~(\ref{eqD02}) (red solid line). The breathing period is plotted with dashed lines for no-noise (cyan), equation chain~(\ref{eqD01}) (grey), two-cumulant model (magenta).
}
  \label{fig4}
\end{figure}

In the noise-free case $\sigma=0$, Abrams {\it et al.} assumed wrapped Cauchy distributions of oscillator phases and found stable chimera states with synchronized subpopulation $\psi_1=\psi_2=\dots=\psi_N=\psi$ and partially synchronized subpopulation $\{\varphi_j\}$ with $0<|Z_1|<1$. This chimera can be either ``stable,'' with constant $|Z_1(t)|$,  or ``breathing,'' with oscillating $|Z_1(t)|$ as one can see in Fig.~\ref{fig3}.
Pikovsky and Rosenblum~\cite{Pikovsky-Rosenblum-2008} shown that for identical oscillators, by virtue of the Watanabe--Strogatz theory~\cite{Watanabe-Strogatz-1993,Watanabe-Strogatz-1994}, these chimera states are neutrally stable with respect to deviations from the Cauchy distribution of phases $\varphi_j$. Mathematically speaking, the Ott--Antonsen manifold $\kappa_1=Z_1$, $\kappa_{m\ge2}=0$ is not transversally attracting. In Fig.~\ref{fig3}b, one can see two quasiperiodic trajectories starting with different initial conditions and maintaining different width of the torus tube (the trajectory segments are shown after the decay of transient oscillations, for $t>2000$).  Oscillation around the Cauchy-distribution solution {\it or} the OA solution derived by Abrams {\it et al.}\ introduces additional frequency into the quasiperiodic chimera dynamics.

The following bifurcation scenario of the OA chimeras was reported in the no-noise case~\cite{Abrams-etal-2008}.
As the coupling asymmetry $A$ increases, the ``stable'' chimera appears via a saddle-node bifurcation at $A_\mathrm{SN}$ (see blue circles in Fig.~\ref{fig4}); at $A_\mathrm{H}$ this chimera becomes unstable via a Hopf bifurcation and the breathing chimera emerges; at $A_\mathrm{HL}$ the cycle of the breathing chimera forms a homoclinic loop and disappears for larger $A$. The latter bifurcation point can be accurately calculated numerically by fitting the divergence law of the cycle period with the asymptotic law $T\approx \tau_1-\tau_2\ln(A_\mathrm{HL}-A)$ known for a homoclinic bifurcation (see dashed lines in Fig.~\ref{fig4}).

The effect of noise on the macroscopic dynamics is threefold: first, it makes the ``stable'' and breathing chimera states found in Ref.~\citen{Abrams-etal-2008} attracting; second, it shifts the bifurcation points; third, the perfect synchrony of the second population is distorted, $|Y_1|=1-\mathcal{O}(\sigma^\alpha)$. Thus, in the presence of noise, the bifurcations of the successors of the chimeras found by Abrams {\it et al.}\ provide a picture of stable macroscopic regimes in the population (unlike to the no-noise case where these chimeras are neutrally stable with respect to the deviations from the Cauchy shape of the distribution of phases).

\begin{table*}[t]
\caption{Bifurcation points of chimera states in hierarchical population~(\ref{eqD01}) for $\vartheta=\pi/2-0.15$ and $\sigma^\alpha=4\times10^{-4}$. The ``exact'' numerical results are calculated with equation chains~(\ref{eqD01}) with 200 terms for each subpopulation; the numerical results for two-cumulant model~(\ref{eqD02}) are marked by ``2CCs''.}
\begin{center}
\begin{tabular}{p{1.8cm}cccccccc}
\hline\hline
 \multicolumn{2}{c}{} & no noise & \multicolumn{2}{c}{$\alpha=2$}
 & \multicolumn{2}{c}{$\alpha=1.5$}
 & \multicolumn{2}{c}{$\alpha=0.5$}
 \\
 & & & exact & 2CCs & exact & 2CCs & exact & 2CCs
 \\
\hline
Saddle-node
 & $A_\mathrm{SN}$ & 0.2505
 & 0.2523 & 0.2523
 & 0.2518 & 0.2517
 & 0.2507 & 0.2509
\\
bifurcation
 & $|Z_1|_\mathrm{SN}$ & 0.7472
 & 0.7517 & 0.7516
 & 0.7500 & 0.7507
 & 0.7509 & 0.7502
\\[5pt]
Hopf bifurca-
 & $A_\mathrm{H}$ & 0.2919
 & 0.2975 & 0.2975
 & 0.2955 & 0.2957
 & 0.2938 & 0.2936
\\
tion
 & $|Z_1|_\mathrm{H}$ & 0.6066
 & 0.6007 & 0.6008
 & 0.6029 & 0.6027
 & 0.6044 & 0.6047
\\[5pt]
Homoclinic bifurcation
 & $A_\mathrm{HL}$ & 0.3239
 & 0.3409 & 0.3417
 & 0.3313 & 0.3331
 & 0.3267 & 0.3246
\\
\hline\hline
\end{tabular}
\end{center}
\label{tab1}
\end{table*}

In Fig.~\ref{fig4} and Table~\ref{tab1}, the results for Gaussian ($\alpha=2$) and non-Gaussian noises are presented. For the Gaussian noise, one can see a noticeable shift of the bifurcation points by noise. Meanwhile, the discrepancy between the results for the two-cumulant model and the `infinite' equation chain is below the accuracy of the calculation of the position of the local bifurcation points (saddle-node and Hopf bifurcations). The results can be distinguished only for the non-local bifurcation of homoclinic loop, where the inaccuracy is also below 5\% of the noise-induced shift of the bifurcation point. In Fig.~\ref{fig4}a, one can visually distinguish only the graphs of the chimera breathing periods for the two models. For the non-Gaussian noises, the noise-induced shift of the bifurcation points is somewhat smaller. The results for the two models can be distinguished in Fig.~\ref{fig4}b,c and Table~\ref{tab1}, but the two-cumulant model reduction still provides a reasonable accuracy of the bifurcation diagrams. Noteworthily, noise significantly widens the parameter domain where the breathing chimera exists; the two-cumulant model reduction gives an adequate account of this widening.

\subsection{Implication for synchronization by common noise}
In Refs.~\citen{Goldobin-Pikovsky-2005,Goldobin-Dolmatova-2019}, the synchronization by common noise in the presence of global coupling is studied for a general class of limit cycle oscillators. In high-synchrony states, the phase fluctuations $\theta_j$ from the synchronous cluster are reported to possess $|\theta|^{-2-2\varepsilon/\sigma^2}$-tails for $\varepsilon>-\sigma^2/2$, where $\mu$ is the coupling coefficient and $\sigma^2$ is the intensity of Gaussian common noise. For moderate repulsive and attractive coupling $-1/2<\varepsilon/\sigma^2<1/2$ these fluctuations generate an $\alpha$-stable noise with $0<\alpha<2$. In this case, closure~(\ref{eq204}) can be beneficial. Moreover, an effective endogenous noise in ensembles of such oscillators with large but finite number of connections~\cite{diVolo-etal-2022} will be also $\alpha$-stable.

\section{Methods}\label{sec:Meth}
\subsection{Delta-correlated non-Gaussian noises}\label{ssec:ASt}
Consider a stochastic dynamical system
\begin{equation}
\dot{\mathbf{x}}=\mathbf{f}(\mathbf{x})+\mathbf{g}(\mathbf{x})\,\xi(t)\,,
\label{eq101}
\end{equation}
where $\mathbf{x}=\{x_1,x_2,...,x_N\}$; the properties of noise $\xi(t)$ will be specified below, generally it is non-Gaussian. We imply the Stratonovich interpretation of this equation.

First, we recall the definition of the characteristic functional for a stochastic process $\xi(t)$~\cite{Klyatskin-1980,Klyatskin-2005}:
\begin{equation}
F^{(\xi)}[v(t)]=\langle e^{i\int_{-\infty}^{+\infty} v(\tau)\,\xi(\tau)\,\mathrm{d}\tau}\rangle,
\label{eq102}
\end{equation}
where $\langle\dots\rangle$ indicates averaging over noise realizations; for some applications one needs
\begin{equation}
F_t^{(\xi)}[v(t)]=\langle e^{i\int^t v(\tau)\,\xi(\tau)\,\mathrm{d}\tau}\rangle.
\label{eq103}
\end{equation}
The logarithm of the characteristic functional
\begin{align}
&\textstyle
\Phi^{(\xi)}[v(t)]\equiv\ln{F^{(\xi)}[v(t)]} \equiv\int\mathrm{d}t_1K_{\xi,1}(t_1)\,iv(t_1)
\nonumber\\
&\textstyle\qquad
+\int\mathrm{d}t_1\int\mathrm{d}t_2K_{\xi,2}(t_1,t_2)\frac{i^2v(t_1)v(t_2)}{2!} \nonumber\\
&\textstyle
+\int\mathrm{d}t_1\int\mathrm{d}t_2\int\mathrm{d}t_3K_{\xi,3}(t_1,t_2,t_3)\frac{i^3v(t_1)v(t_2)v(t_3)}{3!} +\dots\,,
\label{eq104}
\end{align}
where $K_{\xi,n}(t_1,t_2,\dots,t_n)$ is the $n$-th cumulant function; and
\[
\Phi_t^{(\xi)}[v(t)]\equiv\ln{F_t^{(\xi)}[v(t)]}\,.
\]

For a $\delta$-correlated noise, {\em if} cumulant functions can be introduced, they are
\[
K_{\xi,n}(t_1,t_2,\dots,t_n)=\mathcal{K}_{\xi,n}(t_1)\,\delta(t_2-t_1)\dots\delta(t_n-t_1)\,,
\]
where we admit stochastic processes with time-dependent properties, $\mathcal{K}_{\xi,n}(t)$\,. For discrete-time versions of Eq.~(\ref{eq101}) with time stepsize $\Delta t$, one finds $\xi_j=\xi(t_j)$ to be independent random numbers with cumulants $\mathcal{K}_{\xi,n}(t_j)/(\Delta t)^{n-1}$.

If $\mathcal{K}_{\xi,n}(t)$ control the average dynamics of the system, they have to be invariant with respect to $(\Delta t)$, as it tends to 0. Hence, the generated discrete-time signal should possess cumulants $K_{\Delta t,n}\propto(\Delta t)^{-n+1}$, while $K_{\Delta t,n}\propto\sigma^n$, where $\sigma$ is the noise strength. Due to the statistical properties of finite samples (the sums of sets of independent identically distributed random variables), $K_{N\Delta t,n}=NK_{\Delta t,n}$ and, if the discrete-time noise is generated with $K_{\Delta t,n}=\mathcal{K}_{\xi,n}/(\Delta t)^{n-1}$, it holds this property on coarsened time scales. In this respect, it is not inconsistent to generate such a non-Gaussian discrete-time noise.

Further, if we deal with a $\delta$-correlated noise driving dynamical system~(\ref{eq101}), the $\alpha$-stable distributions of $\xi$ are relevant. Indeed, if we adopt for a discrete-time version of Eq.~(\ref{eq101}) a probability density of $\xi$ with tails $\propto|\xi|^{-1-\alpha}$, $0<\alpha<2$, then, by virtue of the generalized central limit theorem, in the limit $\Delta t\to0$, the noise will effectively act as an $\alpha$-stable one (if the tails $\propto|\xi|^{-1-\alpha}$ with $\alpha\ge2$, the noise becomes Gaussian, which is assigned to $\alpha=2$). Thus, in numerical simulations, picking-up an $\alpha$-stable distribution will yield better convergence of the results to the limit $\Delta t\to0$; in theoretical analysis, a physically consistent consideration must employ $\alpha$-stable statistics for a $\delta$-correlated noise.

Now we briefly recall the properties of $\alpha$-stable distributions~\cite{Zolotarev-1986} we will need. The sum of a large number of independent identically distributed random variables, the probability density of which possesses power-law tails $|\xi|^{-1-\alpha}$, is distributed according to the $\alpha$-stable distribution (with the same exponent of the power law for tails) for $0<\alpha<2$ and to the Gaussian distribution ($\alpha=2$) for $\alpha\ge2$. For $\alpha\le0$ the distribution cannot be normalized. Generally, these distributions are specified via their characteristic functions;
\begin{equation}
F(v;\alpha,\beta,c,\mu)=\exp\big(iv\mu-|cv|^\alpha(1+i\beta\mathrm{sign}(v)\Theta)\big)\,,
\label{eq105}
\end{equation}
where $\mu\in\mathbb{R}$ is a shift parameter, $c>0$ is a scale parameter featuring the distribution width, $\beta\in[-1,1]$ called the skewness parameter (typically the term ``skewness'' is used for the third cumulant, which can diverge here),
\[
\Theta=\left\{
\begin{array}{cc}
\tan\left(\frac{\pi\alpha}{2}\right), & \mbox{ for } \alpha\ne1; \\
-\frac{2}{\pi}\ln|v|, & \mbox{ for } \alpha=1.
\end{array}
\right.
\]

It is important to discuss the discrete-time version of a $\delta$-correlated $\alpha$-stable noise. The sum of two independent $\alpha$-stable random variables is an $\alpha$-stable random variable with parameters [see Eq.~(\ref{eq105})] $\mu=\mu_1+\mu_2$, $|c|=(|c_1|^\alpha+|c_2|^\alpha)^{1/\alpha}$, $\beta=(\beta_1|c_1|^\alpha+\beta_2|c_2|^\alpha)/(|c_1|^\alpha+|c_2|^\alpha)$. For the sum of two noise increments $\xi_{\Delta{t}}\Delta{t}$ the equivalent increment for the time step size $(2\Delta{t})$ is $\xi_{2\Delta{t}}2\Delta{t}=2^{1/\alpha}\xi_{\Delta{t}}\Delta{t}$; therefore, $\xi_{\Delta{t}}\propto(\Delta{t})^{1/\alpha-1}$. One must comply with the latter scaling law in numerical simulations.

\subsection{Fractional Fokker--Planck equation}\label{ssec:FFP}
The derivation of the fractional Fokker--Planck equation for an {\em additive} noise is widely presented in the literature and can be readily conducted in the Fourier space as a continuous limit of a pulse noise (e.g., see Ref.~\citen{Toenjes-Pikovsky-2020}). In this section we follow a more involved alternative derivation~\cite{Klyatskin-1980,Klyatskin-2005} for the general case of Eq.~(\ref{eq101}).

For stochastic system~(\ref{eq101}) the probability density function $w(\mathbf{x},t)=\langle\delta(x_1(t)-x_1)\,\delta(x_2(t)-x_2)\dots\delta(x_N(t)-x_N)\rangle =\langle\delta^N(\mathbf{x}(t)-\mathbf{x})\rangle$
obeys the following equation
\begin{align}
\frac{\partial}{\partial t}w(\mathbf{x},t)
+\frac{\partial}{\partial x_j}\big(f_j(\mathbf{x})\,w(\mathbf{x},t)\big)\qquad
\nonumber\\
=\Big\langle\xi(t)\frac{\delta}{\delta\xi(t)}\delta^{N}(\mathbf{x}(t)-\mathbf{x})\Big\rangle\,,
\label{eq106}
\end{align}
where the Einstein convention of summation over repetitive indices is adopted and $\frac{\delta}{\delta\xi(t)}$ is a variational derivative for $\delta^{N}(\mathbf{x}(t)-\mathbf{x})$, which is a functional of $\xi(\tau)|_{\tau\in(-\infty,t]}$. For a delta-correlated process $\xi(t)$, the time-derivative of the logarithm of characteristic functional
\[
\dot\Phi_t[v(t)]=\frac{\langle i\xi(t)v(t)e^{i\int^t\xi(\tau)v(\tau)\mathrm{d}\tau}\rangle} {\langle e^{i\int^t\xi(\tau)v(\tau)\mathrm{d}\tau}\rangle}
\]
at $t$ involves only $\xi(t)$, as averaging of $\langle e^{i\int^t\xi(\tau)v(\tau)\mathrm{d}\tau}\rangle$ can be factorized for averagings over $\xi(t)$ and realization $\xi(\tau)|_{\tau\in(-\infty,t-0)}$.
For a discrete time, $\xi(t)\mathrm{d}t\propto(\mathrm{d}t)^{1/\alpha}$ and, after cancelling the common part $\langle e^{i\int^{t-\mathrm{d}t}\xi(\tau)v(\tau)\mathrm{d}\tau}\rangle$, the average in denominator is $\sim\big(1+\mathcal{O}(v^2(\mathrm{d}t)^{2/\alpha})\big)$, while in numerator one finds $\sim v^2(\mathrm{d}t)^{1/\alpha+1}$. The latter can be finite where the former turns to 1. Hence, in Eq.~(\ref{eq106}),
\[
\Big\langle\xi(t)\frac{\delta}{\delta\xi(t)}\delta^{N}(\mathbf{x}(t)-\mathbf{x})\Big\rangle =\Big\langle\dot\Phi_t^{(\xi)}\!\!\left[\frac{1}{i}\frac{\delta}{\delta\xi(t)}\right] \delta^{N}(\mathbf{x}(t)-\mathbf{x})\Big\rangle.
\]
Employing $\frac{\delta}{\delta\xi(t)} =\frac{\delta\mathbf{x}(t)}{\delta\xi(t)}\cdot\frac{\delta}{\delta\mathbf{x}(t)}$, $\frac{\delta\mathbf{x}(t)}{\delta\xi(t)} =\frac{\delta}{\delta\xi(t)} \int^t\big(\mathbf{f}(\mathbf{x}(\tau))+\xi(\tau)\,\mathbf{g}(\mathbf{x}(\tau))\big)\mathrm{d}\tau =\mathbf{g}(\mathbf{x}(t))$, and $\frac{\delta}{\delta\mathbf{x}(t)}\delta^{N}(\mathbf{x}(t)-\mathbf{x})=-\frac{\delta}{\delta\mathbf{x}}\delta^{N}(\mathbf{x}(t)-\mathbf{x})$, one obtains
\begin{align}
\Big\langle\dot\Phi_t^{(\xi)}\!\!\left[\frac{1}{i}\frac{\delta}{\delta\xi(t)}\right] \delta^{N}(\mathbf{x}(t)-\mathbf{x})\Big\rangle &=\langle\dot\Phi_t^{(\xi)}[i\hat{Q}]\delta^{N}(\mathbf{x}(t)-\mathbf{x})\rangle
\nonumber\\
&=\dot\Phi_t^{(\xi)}[i\hat{Q}]w(\mathbf{x},t)\,,
\nonumber
\end{align}
where $\hat{Q}(\cdot)\equiv\frac{\partial}{\partial x_j}\left(g_j(\mathbf{x})(\cdot)\right)$.

Thus, we arrive to the Klyatskin's result~\cite{Klyatskin-1980,Klyatskin-2005}:
\begin{equation}
\frac{\partial w}{\partial t}+
\frac{\partial}{\partial x_j}\big(f_j(\mathbf{x})w\big)
-\dot\Phi_t^{(\xi)}(i\hat{Q})w=0\,,
\label{eq107}
\end{equation}
or, {\em if} the representation by a cumulant function series is possible,
\begin{equation}
\frac{\partial w}{\partial t}+
\frac{\partial}{\partial x_j}\big(f_j(\mathbf{x})w\big)
-\sum_{n=1}^{\infty} \frac{\mathcal{K}_{\xi,n}}{n!}(-\hat{Q})^nw=0\,.
\label{eq108}
\end{equation}

For an important case of sign-symmetric noise $\xi(t)$ with $\alpha$-stable distribution,
one sets $\mu=0$ and $\beta=0$ in (\ref{eq105}); therefore,
$\Phi_t^{(\xi)}[v(t)]=-\int^t\ln{F(v(\tau);\alpha,\beta=0,c,\mu=0)} \mathrm{d}\tau
=-\int^t|cv(\tau)|^\alpha\mathrm{d}\tau$ and its time-derivative
\[
\dot\Phi_t^{(\xi)}[v(t)] =-|cv(t)|^\alpha.
\]
For the noise strength $c=\sigma$, one can recast Eq.~(\ref{eq107}) as
\begin{equation}
\frac{\partial w}{\partial t}+
\frac{\partial}{\partial x_j}\big(f_j(\mathbf{x})w\big)
-\dot\Phi_t^{(\xi)}(i\sigma\hat{Q};c=1)w=0\,,
\label{eq109}
\end{equation}
where ``$c=1$'' explicitly indicates how the noise amplitude interacts with operator $\hat{Q}$: one can take the characteristic functional with $c=1$ and multiply its argument by $\sigma$.

Here and hereafter, we restrict our consideration to the case of a phase equation with additive noise, $g(\varphi)=1$, and $f(\varphi,t)=\omega(t)+\mathrm{Im}(h(t)e^{-i\varphi})$\,:
\begin{equation}
\dot{\varphi}=\omega(t)+\mathrm{Im}(h(t)e^{-i\varphi})+\sigma\xi(t)\,.
\label{eq110}
\end{equation}
In this case, Eq.~(\ref{eq109}) reads
\begin{equation}
\frac{\partial w}{\partial t}+
\frac{\partial}{\partial\varphi}\big(f(\varphi)w\big)
-\dot\Phi_t^{(\xi)}\left(i\sigma\frac{\partial}{\partial\varphi}\right)w=0\,.
\label{eq111}
\end{equation}
In Fourier space, $w(\varphi,t)=(2\pi)^{-1}\sum_{m=-\infty}^{+\infty}a_m e^{-im\varphi}$ with $a_0=1$ and $a_{-m}=a_m^\ast$, and
\begin{equation}
\dot{a}_m=im\omega a_m+mh\,a_{m-1}-mh^\ast a_{m+1} +\dot\Phi_t^{(\xi)}(\sigma m)\,a_m\,.
\label{eq112}
\end{equation}
As the equations for $a_{m>1}$ do not involve $a_{m<0}$ and $a_0=1$, function $\dot\Phi_t^{(\xi)}(\sigma m)$ acquires analytic form $\dot\Phi_t^{(\xi)}(\sigma m)=-(\sigma m)^\alpha$, which means that it is well defined for an operator in Eq.~(\ref{eq107}).

Notice, without the restriction $m>0$, the operator $\dot\Phi_t^{(\xi)}(i\sigma\frac{\partial}{\partial\varphi}) =-\sigma^\alpha|\frac{\partial}{\partial\varphi}|^\alpha$ is nonanalytic with respect to $m$; it involves the Riesz fractional derivative often~\cite{Chechkin-etal-2003,Toenjes-etal-2013} written as $\frac{\partial^\alpha}{\partial|\varphi|^\alpha}$.

To summarize, for a $\delta$-correlated $\alpha$-stable symmetric additive noise $\xi(t)$ of strength $\sigma$, the probability density function is governed by the following equation for Fourier modes:
\begin{equation}
\dot{a}_m=im\omega a_m+mh\,a_{m-1}-mh^\ast a_{m+1} -(\sigma m)^\alpha a_m\,.
\label{eq113b}
\end{equation}

\subsection{Circular cumulant equation chain}\label{ssec:CCChain}
Normally, the infinite chain of circular cumulant equations can be derived in a regular way via the dynamics of the moment-generating function~(\ref{eqFkt}) of the cyclic variable $\varphi$. Following the conventional procedure~\cite{Tyulkina-etal-2018,Goldobin-Dolmatova-2020,Zheng-Kotani-Jimbo-2021,Goldobin-2021}, one calculates the time-derivative
\begin{align}
&\frac{\partial}{\partial t}F(k,t)=\sum_{m=0}^{\infty}\dot{Z}_m(t)\frac{k^m}{m!}
\nonumber\\
&=(i\omega_0-\gamma)k\frac{\partial}{\partial k} F+hkF-h^\ast k\frac{\partial^2}{\partial k^2}F +\dot\Phi_t^{(\xi)}\left(\sigma k\frac{\partial}{\partial k}\right)F\,,
\nonumber
\end{align}
where $\dot{Z}_m$ are substituted from Eq.~(\ref{eq114}). With
$\Psi(k,t)=\frac{k}{F}\frac{\partial F}{\partial k}
=k\frac{\partial}{\partial k}\ln{F(k,t)}$ (\ref{eqPSIkt}), one can write $\dot{\Psi}=k\frac{\partial}{\partial k}\frac{\dot{F}}{F}$; therefore,
\begin{align}
\frac{\partial\Psi}{\partial t}&=(i\omega_0-\gamma)k\frac{\partial}{\partial k}\left(\frac{k}{F}\frac{\partial F}{\partial k}\right)
+hk-h^\ast k\frac{\partial}{\partial k}\left(\frac{k}{F}\frac{\partial^2F}{\partial k^2}\right)
\nonumber\\
&\qquad\qquad
{}+k\frac{\partial}{\partial k}\left(
 \frac{1}{F}\dot\Phi_t^{(\xi)}\left(\sigma k\frac{\partial}{\partial k}\right)F\right)
\nonumber\\
&=(i\omega_0-\gamma)k\frac{\partial}{\partial k}\Psi+hk
 -h^\ast k\frac{\partial}{\partial k}
 \left(k\frac{\partial}{\partial k}\frac{\Psi}{k}+\frac{\Psi^2}{k}\right)
\nonumber\\
&\qquad\qquad
{}-\sigma^\alpha k\frac{\partial}{\partial k}\left(
 \frac{1}{F}\left(k\frac{\partial}{\partial k}\right)^\alpha F\right)\,.
\label{eqap01}
\end{align}
Substituting power series~(\ref{eqPSIkt}) into (\ref{eqap01}) and collecting terms with $k^m$, one obtains
\begin{align}
\dot\kappa_m&=m(i\omega_0-\gamma)\kappa_m+h\delta_{m\,1}
\nonumber\\
&\quad{}
 -h^\ast\Big(m^2\kappa_{m+1}+m\sum_{n=1}^{m}\kappa_{m-n+1}\kappa_{n}\Big)
 -m\sigma^\alpha\mathcal{G}_m^{(\alpha)}\,,
\label{eqap02}
\end{align}
where $\mathcal{G}_m^{(\alpha)}$ are the coefficients of the power series
\[
\mathcal{G}^{(\alpha)}\equiv\frac{1}{F}\left(k\frac{\partial}{\partial k}\right)^\alpha F
=\frac{1}{F}\left(k\frac{\partial}{\partial k}\right)^{\alpha-1}(F\Psi)
\equiv\sum_{m=1}^{\infty}\mathcal{G}_m^{(\alpha)}k^m.
\]

For integer $\alpha$, one can derive explicit expressions of $\mathcal{G}_m^{(\alpha)}$ for all $m$. For $\alpha=1$, the series $\mathcal{G}^{(1)}=\Psi$ and the effect of noise can be effectively represented by a plain shift $\gamma\to\gamma+\sigma$; for $\alpha=2$,
the series $\mathcal{G}^{(2)}=\frac{1}{F}k\frac{\partial}{\partial k}(F\Psi)
=k\frac{\partial\Psi}{\partial k}+\Psi^2$ and
\begin{equation}
\mathcal{G}_m^{(2)}=m\kappa_m+\sum_{n=1}^{m-1}\kappa_{m-n}\kappa_{n}\,.
\label{eqapG2m}
\end{equation}
For fractional $\alpha$, the coefficients of the series
\[
\mathcal{G}^{(\alpha)}=e^{-\sum_{n=1}^{\infty}\kappa_n\frac{k^n}{n}}
\left(k\frac{\partial}{\partial k}\right)^\alpha
e^{\sum_{l=1}^{\infty}\kappa_l\frac{k^l}{l}},
\]
where the fractional derivative does not change the exponent of $k^m$ but creates the multiplier $m^\alpha$ for the $k^m$-term of the right-hand-side exponential, cannot be represented by an explicit formula.

The recursive procedure for calculating $\mathcal{G}_m^{(\alpha)}$ can be useful. Using recursive direct and inverted formulae for circular cumulants and moments
\begin{align}
\kappa_m&=\frac{Z_m}{(m-1)!}-\sum_{n=1}^{m-1}\frac{\kappa_nZ_{m-n}}{(m-n)!}\,,
\label{eqap03}
\\
Z_m&=(m-1)!\kappa_m+\sum_{n=1}^{m-1}\frac{(m-1)!}{(m-n)!}\kappa_nZ_{m-n}\,,
\label{eqap04}
\end{align}
one can write the time-derivative of (\ref{eqap03})
\[
\dot\kappa_m=\frac{\dot{Z}_m}{(m-1)!}-\sum_{n=1}^{m-1}\frac{\kappa_n\dot{Z}_{m-n}+\dot\kappa_nZ_{m-n}}{(m-n)!}
\]
and find
\begin{equation}
m\mathcal{G}_m^{(\alpha)}=\frac{m^\alpha Z_m}{(m-1)!}
-\sum_{n=1}^{m-1}\frac{\big(\kappa_n(m-n)^\alpha+n\mathcal{G}_n^{(\alpha)}\big)Z_{m-n}}{(m-n)!}\,.
\label{eqap05}
\end{equation}
Starting from $(Z_1,\mathcal{G}_1^{(\alpha)})=(\kappa_1,\kappa_1)$, one can recursively employ Eqs.~(\ref{eqap05}) and (\ref{eqap04}) to calculate $\mathcal{G}_m^{(\alpha)}$ as a function of $\kappa_n$ with $n=1,...,m$ for all $m$.

Employing the recursive procedure, one obtains
\begin{subequations}
\label{eqap06}
\begin{align}
\mathcal{G}_1^{(\alpha)}&=\kappa_1\,,
\label{eqap06a}
\\
\mathcal{G}_2^{(\alpha)}&=2^{\alpha-1}\kappa_2+(2^{\alpha-1}-1)\kappa_1^2\,,
\label{eqap06b}
\\
\mathcal{G}_3^{(\alpha)}&=3^{\alpha-1}\kappa_3+\frac{3^\alpha-2^\alpha-1}{2}\kappa_2\kappa_1 +\frac{3^{\alpha-1}-2^\alpha+1}{2}\kappa_1^3\,,
\label{eqap06c}
\\
\mathcal{G}_4^{(\alpha)}&=4^{\alpha-1}\kappa_4+\frac{4^\alpha-3^\alpha-1}{3}\kappa_3\kappa_1 +\frac{4^{\alpha-1}-2^{\alpha-1}}{2}\kappa_2^2
\nonumber\\
&\hspace{-10pt}
{}+\frac{4^\alpha-2\times3^\alpha+2}{4}\kappa_2\kappa_1^2 +\frac{4^{\alpha-1}-3^\alpha+3\times2^{\alpha-1}-1}{6}\kappa_1^4\,,
\label{eqap06d}
\\
\mathcal{G}_5^{(\alpha)}&=5^{\alpha-1}\kappa_5+\frac{5^\alpha-4^\alpha-1}{4}\kappa_4\kappa_1 +\frac{5^\alpha-3^\alpha-2^\alpha}{6}\kappa_3\kappa_2
\nonumber\\
&
{}+\frac{5^\alpha-2\times4^\alpha+3^\alpha-2^\alpha+2}{6}\kappa_3\kappa_1^2
\nonumber\\
&
{}+\frac{5^\alpha-4^\alpha-2\times3^\alpha+2^{\alpha+1}+1}{8}\kappa_2^2\kappa_1
\nonumber\\
&
{}+\frac{5^\alpha-3\times4^\alpha+2\times3^\alpha+2^{\alpha+1}-3}{12}\kappa_2\kappa_1^3
\nonumber\\
&
{}+\frac{5^{\alpha-1}-4^\alpha+2\times3^\alpha-2^{\alpha+1}+1}{24}\kappa_1^5\,,
\label{eqap06e}
\\
&\dots\;.
\nonumber
\end{align}
\end{subequations}
With Eqs.~(\ref{eqap06}) one can see that all the terms but the first ones vanish for $\alpha=1$. For $\alpha=2$, the first terms turn into $m\kappa_m$, which is the first term of (\ref{eqapG2m}), the coefficients of biproducts and squares simplify to $2$ and $1$, respectively, which yields the sum in (\ref{eqapG2m}), and the coefficients of all the terms with more than 2 cumulants turn to $0$. For a fractional $\alpha$, $0<\alpha<2$, all the coefficients in (\ref{eqap06}) (for the first five circular cumulant equations) are nonzero and the expressions keep a sophisticated form.

\section{Conclusion}\label{sec:concl}
Within the framework of the circular cumulant approach, one can construct a reasonably accurate macroscopic description of the dynamics of phase oscillators (elements) subject to non-Gaussian $\delta$-correlated noises. By virtue of the generalized central limit theorem, for $\delta$-correlated noises, only the case of $\alpha$-stable noises is physically meaningful; any other non-Gaussian noise will converge to the $\alpha$-stable one as we switch from a discrete time to the continuous time limit $\Delta t\to0$.

The variables of our low dimensional model reduction are associated with the leading Fourier modes of the probability density distribution. Therefore, within the framework of our approach we circumvent the issue of unphysical behavior of the probability density (formation of discontinuities, negativity, etc.), which is typical for Fokker--Planck-type equations for non-Gaussian noise with truncated cumulant series.

In the literature, the earlier low dimensional models of this kind were basically one-circular-cumulant reductions. For the Ott--Antonsen Ansatz~\cite{Ott-Antonsen-2008,Ott-Antonsen-2009} corresponding to the wrapped Cauchy distributions of phases, one substitutes $Z_m=Z_1^m$ into the first equation of infinite chain~(\ref{eq114}) in terms of circular moments, or $\kappa_{m\ge2}=0$ into the first equation of~(\ref{eqap02}) in terms of circular cumulants. For the alternative option of the wrapped Gaussian distribution approximation~\cite{Zaks-etal-2003,Sonnenschein-etal-2013,Sonnenschein-etal-2015,Hannay-etal-2018}, one substitutes $Z_m=|Z_1|^{m^2-m}Z_1^m$ or $\kappa_2=(|\kappa_1|^2-1)\kappa_1^2$, $\kappa_3=(0.5|\kappa_1|^6-1.5|\kappa_1|^2+1)\kappa_1^3$, etc.~\cite{Goldobin-etal-2018,Goldobin-Dolmatova-2019b} into the first equation of the corresponding infinite chain. In the Gaussian case, the researchers~\cite{Zaks-etal-2003,Sonnenschein-etal-2013,Sonnenschein-etal-2015} often switch to the interpretation in terms of a phase treated like a variable on the infinite line and speak of the center phase of the distribution and the variance of phase deviations as of the first and second cumulants. However, these two real-valued {\it linear-variable} quantifiers correspond merely to the first {\it circular} moment or cumulant~\cite{Goldobin-Dolmatova-2019b}, which is complex-valued:  $|Z_1|=e^{-\mathrm{Var}/2}$ and $\arg{Z_1}$ is the distribution center. The one-cumulant reductions strip the system from the freedom of self-adjustment of the distribution shape and type. Meanwhile, the two-cumulant representation can be useful for tracking the diversity of the distribution types and even the loss of unimodality~\cite{Tyulkina-etal-2019}.

The implementation of the approach is demonstrated for the case of the Kuramoto ensemble with non-Gaussian noise. The results of direct numerical simulation for the ensemble of $N=1500$ phases oscillators~(\ref{eq201}), the numerical solution of the chain of equations~(\ref{eq114}) for the Kuramoto--Daido order parameters $Z_m=N^{-1}\sum_{j=1}^{N}e^{im\varphi_j}$ in the thermodynamic limit $N\to\infty$ (Fourier modes of the probability density) with 200 elements $Z_m$, and the analytical solutions of the two-cumulant approximations~(\ref{eq202}) and (\ref{eq207}) are in good agreement with each other.

The two-cumulant model reduction was also tested for an essentially non-stationary situation. The model was used to survey the noise effect on the bifurcations of chimera states in a hierarchical population of coupled oscillators and the results were compared against the background of those simulated with the Fourier expansion. For saddle-node and Hopf bifurcations, the agreement is excellent; notice, in terms of the numerically simulated variables these are a saddle-node bifurcation of a cycle and a Hopf bifurcation of a Poincar\'e map, i.e.\ the branching of an invariant torus from a cycle. For a homoclinic bifurcation, which is nonlocal, the quantitative agreement is worse than for the two local bifurcations, but still reasonable (Table~\ref{tab1}).

The chimera example is outspoken for our cause, since the series of the Fourier coefficients $Y_m$ of the synchronous subpopulation does not decay for vanishing noise intensity. In the absence of noise, the Ott--Antonsen solution of the synchronized subpopulation is attracting~\cite{Tyulkina-etal-2019} and the OA Ansatz $Y_m=Y_1^m$ allows one to circumvent the absence of the decay of $|Y_m|$. A weak noise first makes the synchrony imperfect, $|Y_1|=1-\mathcal{O}(\sigma^\alpha)$, and second it breaks the applicability of the OA Ansatz, necessitating the numerical simulation of the chain of equations for $Y_m$. For $\alpha=1$, the decay of $Y_m$ is still exactly the OA one and for, say $m=200$, $|Y_{200}|=|Y_1|^{200}\sim e^{-200\sigma}$, which is still nonsmall even for the values of $\sigma$ where all chimera regimes are already suppressed. For $\alpha>1$, the noise creates a faster decay of $|Y_m|$; $\ln|Y_m|\propto -(\sigma m)^\alpha$ and $\exp(-200^\alpha\sigma^\alpha)$ for $\sigma^\alpha\sim10^{-3}$ is small enough for one to safely adopt the truncation $Y_{201}=0$ for the direct numerical simulation of the equation chain. For $\alpha<1$, the noise does not enhance the OA decay of $|Y_m|\sim|Y_1|^m$; therefore, the employment of the truncated equation chain for $Y_m$ is most challenging. For any $\alpha$, one cannot use the Fourier representation for studying the limiting case of $\sigma^\alpha\to0$, but the limitations on the reachable values of $\sigma^\alpha$ are more severe for $\alpha<1$. Simultaneously, the two-cumulant model reduction works well for the limiting case of $\sigma^\alpha\to0$.

For the parameter range examined in this paper (Sec.~\ref{ssec:Abr}), the noise only shifts the bifurcation points and makes the Abrams--Mirollo--Strogatz--Wiley chimeras attracting. However, for other values of coupling parameters $(\vartheta,A)$ or a stronger noise, new types of chimeras emerge. Moreover, some of these new chimera types emerge via the bifurcations of the nearly-synchronous subpopulation, which can be only partly studied in terms of $Y_m$-series, while the two-cumulant reduction works uniformly well. The full picture of the chimera regimes and their bifurcations is too diverse and complex and will be the subject of further studies, but it illustrates the usefulness of the two-cumulant model reduction.

\section*{ACKNOWLEDGMENTS}
The authors are thankful to L.~S.\ Klimenko and  A.\ Pikovsky for fruitful discussions and comments and acknowledge the financial support from RSF (Grant no.\ 23-12-00180).

\section*{AUTHOR DECLARATIONS}
\subsection*{Conflict of Interest}
The authors have no conflicts to disclose.

\subsection*{Author Contributions}
\textbf{Anastasiya V.\ Dolmatova:}
Conceptualization (supporting);
Formal analysis (equal);
Investigation (equal);
Software (equal);
Writing -- original draft (supporting).
\textbf{Irina V.\ Tyulkina:}
Conceptualization (supporting);
Investigation (supporting);
Software (equal);
Writing -- original draft (supporting).
\textbf{Denis S.\ Goldobin:}
Conceptualization (lead);
Formal analysis (equal);
Investigation (equal);
Software (equal);
Writing -- original draft (lead).

\section*{DATA AVAILABILITY}
The data that support the findings of this study are available within the article and shown in the figures.








\begin{thebibliography}{20}

\bibitem{Mandelbrot-1963}
 B. Mandelbrot,
 ``New methods in statistical economics,''
 The Journal of Political Economy {\bf 71}(5), 421--440 (1963).


\bibitem{Chechkin-etal-2003}
 A.~V.\ Chechkin, J.\ Klafter, V.~Yu.\ Gonchar, R.\ Metzler, and L.~V.\ Tanatarov,
 ``Bifurcation, bimodality, and finite variance in confined L\'evy flights,''
 Phys.\ Rev.\ E {\bf 67}(1), 010102(R) (2003).\\
 https://doi.org/10.1103/PhysRevE.67.010102

\bibitem{Toenjes-etal-2013}
 R.\ Toenjes, I.~M.\ Sokolov, and E.~B.\ Postnikov,
 ``Nonspectral Relaxation in One Dimensional Ornstein-Uhlenbeck Processes,''
 Phys.\ Rev.\ Lett. {\bf 110}(15), 150602 (2013).
 https://doi.org/10.1103/PhysRevLett.110.150602


\bibitem{Toenjes-Pikovsky-2020}
 R.\ T\"onjes and A.\ Pikovsky,
 ``Low-dimensional description for ensembles of identical phase oscillators subject to Cauchy noise,''
 Phys.\ Rev.\ E {\bf 102}(5), 052315 (2022).
 https://doi.org/10.1103/PhysRevE.102.052315



\bibitem{Pietras-etal-2023}
 B.\ Pietras, R.\ Cestnik, and A.\ Pikovsky,
 ``Exact finite-dimensional description for networks of globally coupled spiking neurons,''
 Phys.\ Rev.\ E {\bf 107}(2), 024315 (2023).
 https://doi.org/10.1103/PhysRevE.107.024315


\bibitem{Pyragas2-2023}
 V.\ Pyragas and K.\ Pyragas,
 ``Effect of Cauchy noise on a network of quadratic integrate-and-fire neurons with non-Cauchy heterogeneities,''
 Phys.\ Lett.\ A {\bf 480}, 128972 (2023).\\
 https://doi.org/10.1016/j.physleta.2023.128972

\bibitem{Kostin-etal-2023}
 V.~A.\ Kostin, V.~O.\ Munyaev, G.~V.\ Osipov, and L.~A.\ Smirnov,
 ``Synchronization transitions and sensitivity to asymmetry in the bimodal Kuramoto systems with Cauchy noise,''
 unpublished (2023).\\
 https://doi.org/10.48550/arXiv.2212.05858

\bibitem{Goldobin-Pikovsky-2005}
 D.~S.\ Goldobin and A.\ Pikovsky,
 ``Synchronization and desynchronization of self-sustained oscillators by common noise,''
 Phys.\ Rev.\ E {\bf 71}(4), 045201(R) (2005).
 https://doi.org/10.1103/PhysRevE.71.045201


\bibitem{Goldobin-Dolmatova-2019}
 D.~S.\ Goldobin and A.~V.\ Dolmatova,
 ``Interplay of the mechanisms of synchronization by common noise and global coupling for a general class of limit-cycle oscillators,''
 Commun.\ Nonlinear Sci.\ Numer.\ Simulat. {\bf 75}, 94--108 (2019).
 https://doi.org/10.1016/j.cnsns.2019.03.026


\bibitem{Ott-Antonsen-2008}
 E.\ Ott and T.~M.\ Antonsen,
 ``Low dimensional behavior of large systems of globally coupled oscillators,''
 Chaos {\bf 18}(3), 037113 (2008).\\
 https://doi.org/10.1063/1.2930766


\bibitem{Ott-Antonsen-2009}
 E.\ Ott and T.~M.\ Antonsen,
 ``Long time evolution of phase oscillator systems,''
 Chaos {\bf 19}(2), 023117 (2009).
 https://doi.org/10.1063/1.3136851


\bibitem{Tyulkina-etal-2018}
 I.~V. Tyulkina, D.~S. Goldobin, L.~S. Klimenko, and A. Pikovsky,
 ``Dynamics of noisy oscillator populations beyond the Ott-Antonsen ansatz,''
 Phys.\ Rev.\ Lett. {\bf 120}(26), 264101 (2018).\\
 https://doi.org/10.1103/PhysRevLett.120.264101


\bibitem{Goldobin-etal-2018}
 D.~S. Goldobin, I.~V. Tyulkina, L.~S. Klimenko, and A. Pikovsky,
 ``Collective mode reductions for populations of coupled noisy oscillators,''
 Chaos {\bf 28}(10), 101101 (2018).
 https://doi.org/10.1063/1.5053576


\bibitem{Goldobin-2019}
 D.~S.\ Goldobin,
 ``Relationships between the Distribution of Watanabe-Strogatz Variables and Circular Cumulants for Ensembles of Phase Elements,''
 Fluct.\ Noise Lett. {\bf 18}(2), 1940002 (2019).\\
 https://doi.org/10.1142/S0219477519400029


\bibitem{Goldobin-Dolmatova-2019b}
 D.~S. Goldobin and A.~V. Dolmatova,
 ``Ott-Antonsen ansatz truncation of a circular cumulant series,''
 Phys.\ Rev.\ Research {\bf 1}(3), 033139 (2019).\\
 https://doi.org/10.1103/PhysRevResearch.1.033139


\bibitem{Goldobin-Dolmatova-2020}
 D.~S. Goldobin and A.~V. Dolmatova,
 ``Circular cumulant reductions for macroscopic dynamics of Kuramoto ensemble with multiplicative intrinsic noise,''
 J.\ Phys.\ A: Math.\ Theor. {\bf 53}(8), 08LT01 (2020).\\
 https://doi.org/10.1088/1751-8121/ab6b90


\bibitem{Ratas-Pyragas-2019}
 I.\ Ratas and K.\ Pyragas,
 ``Noise-induced macroscopic oscillations in a network of synaptically coupled quadratic integrate-and-fire neurons,''
 Phys.\ Rev.\ E {\bf 100}(5), 052211 (2019).\\
 https://doi.org/10.1103/PhysRevE.100.052211


\bibitem{Zheng-Kotani-Jimbo-2021}
 T.\ Zheng, K.\ Kotani, and Y.\ Jimbo,
 ``Distinct effects of heterogeneity and noise on gamma oscillation in a model of neuronal network with different reversal potential,''
 Sci.\ Rep. {\bf 11}(1), 12960 (2021).\\
 https://doi.org/10.1038/s41598-021-91389-8


\bibitem{Goldobin-2021}
 D.~S. Goldobin,
 ``Mean-field models of populations of quadratic integrate-and-fire neurons with noise on the basis of the circular cumulant approach,''
 Chaos {\bf 31}(8), 083112 (2021).
 https://doi.org/10.1063/5.0061575


\bibitem{diVolo-etal-2022}
 M. di Volo, M. Segneri, D.~S. Goldobin, A. Politi, and A. Torcini,
 ``Coherent oscillations in balanced neural networks driven by endogenous fluctuations,''
 Chaos {\bf 32}(2), 023120 (2022).
 https://doi.org/10.1063/5.0075751


\bibitem{Kuramoto-1975}
 Y.\ Kuramoto,
 in
 {\em International Symposium on Mathematical Problems in Theoretical Physics},
  edited by H.\ Araki,
 Springer Lecture Notes Phys., Vol.\ 39 (Springer, New York, 1975), p.\ 420.


\bibitem{Kuramoto-1984}
 Y.\ Kuramoto,
 {\em Chemical Oscillations, Waves and Turbulence}
 (Springer, Berlin, 1984).
 https://doi.org/10.1007/978-3-642-69689-3


\bibitem{Chambers-etal-1976}
 J.~M.\ Chambers, C.~L.\ Mallows, and B.~W.\ Stuck,
 ``A Method for Simulating Stable Random Variables,''
 J.\ Am.\ Stat.\ Assoc. {\bf 71}(354), 340--344 (1976).


\bibitem{Zolotarev-1986}
 V.~M.\ Zolotarev,
 {\em One-dimensional stable distribution}, Translations of Mathematical Monographs, vol.\ 65 (American Mathematical Society, Providence, 1986).


\bibitem{Misiorek-Weron-2012}
 A.\ Misiorek and R.\ Weron,
 {\em Heavy-Tailed Distributions in VaR Calculations}
 in {\em Springer Handbooks of Computational Statistics} edited by J.~E.\ Gentle, W.~K.\ H\"ardle, and Yu.\ Mori
 (Springer, Berlin Heidelberg, 2012), pp. 1025--1059.
 https://doi.org/10.1007/978-3-642-21551-3\_34


\bibitem{Abrams-etal-2008}
 D.~M.\ Abrams, R.\ Mirollo, S.~H.\ Strogatz, and D.~A.\ Wiley,
 ``Solvable Model for Chimera States of Coupled Oscillators,''
 Phys.\ Rev.\ Lett.\ {\bf 101}, 084103 (2008).
 https://doi.org/10.1103/PhysRevLett.101.084103

\bibitem{Pikovsky-Rosenblum-2008}
 A.\ Pikovsky and M.\ Rosenblum,
 ``Partially Integrable Dynamics of Hierarchical Populations of Coupled Oscillators,''
 Phys.\ Rev.\ Lett. {\bf 101}, 264103 (2008).
 https://doi.org/10.1103/PhysRevLett.101.264103

\bibitem{Watanabe-Strogatz-1993}
 S.\ Watanabe and S.~H.\ Strogatz,
 ``Integrability of a globally coupled oscillator array,''
 Phys.\ Rev.\ Lett. {\bf 70}(16), 2391--2394 (1993).\\
 https://doi.org/10.1103/PhysRevLett.70.2391

\bibitem{Watanabe-Strogatz-1994}
 S.\ Watanabe and S.~H.\ Strogatz,
 ``Constants of motion for superconducting Josephson arrays,''
 Phys.\ D {\bf 74}(3--4), 197--253 (1994).\\
 https://doi.org/10.1016/0167-2789(94)90196-1


\bibitem{Klyatskin-1980}
 V.~I.\ Klyatskin,
 {\em Stokhasticheskiye uravneniya i volny v sluchayno-neodnorodnykh sredakh} [in Russian, {\em Stochastic equations and waves in randomly inhomogeneous media}]
 (Nauka, Moscow, 1980) 337~p.


\bibitem{Klyatskin-2005}
 V.\ Klyatskin,
 {\em Dynamics of Stochastic Systems},
 1st ed.\ (Elsevier Science, 2005) 212~p.
 https://doi.org/10.1016/B978-0-444-51796-8.X5000-9


\bibitem{Zaks-etal-2003}
 M.~A.\ Zaks, A.~B.\ Neiman, S.\ Feistel, and L.\ Schimansky-Geier,
 ``Noisecontrolled oscillations and their bifurcations in coupled phase oscillators,''
 Phys.\ Rev.\ E {\bf 68}(6), 066206 (2003).\\
 https://doi.org/10.1103/PhysRevE.68.066206

\bibitem{Sonnenschein-etal-2013}
 B.\ Sonnenschein and L.\ Schimansky-Geier,
 ``Approximate solution to the stochastic Kuramoto model,''
 Phys.\ Rev.\ E {\bf 88}(5), 052111 (2013).\\
 https://doi.org/10.1103/PhysRevE.88.052111

\bibitem{Sonnenschein-etal-2015}
 B.\ Sonnenschein, Th.~K.~D.~M.\ Peron, F.~A.\ Rodrigues, J.\ Kurths, and L.\ Schimansky-Geier, ``Collective dynamics in two populations of noisy oscillators with asymmetric interactions,''
 Phys.\ Rev.\ E {\bf 91}(6), 062910 (2015).\\
 https://doi.org/10.1103/PhysRevE.91.062910

\bibitem{Hannay-etal-2018}
 K.~M.\ Hannay, D.~B.\ Forger, and V.\ Booth,
 ``Macroscopic models for networks of coupled biological oscillators,''
 Sci.\ Adv. {\bf 4}(8), e1701047 (2018).\\
 https://doi.org/10.1126/sciadv.1701047

\bibitem{Tyulkina-etal-2019}
 I.~V.\ Tyulkina, D.~S.\ Goldobin, L.~S.\ Klimenko, and A.\ Pikovsky,
 ``Two-bunch solutions for the dynamics of Ott–Antonsen phase ensembles,''
 Radiophys.\ Quantum Electron.\ {\bf 61}(8--9), 640--649 (2019).\\
 http://doi.org/10.1007/s11141-019-09924-7



\end{thebibliography}
\end{document}